\documentclass[11pt]{article}

\usepackage[margin=1in]{geometry}
\usepackage{amsmath,amssymb}
\usepackage{graphicx}
\usepackage{booktabs}
\usepackage{tabularx}
\usepackage{multirow}
\usepackage{array}
\usepackage{xcolor}
\usepackage[colorlinks=true,citecolor=blue,linkcolor=blue,urlcolor=blue]{hyperref}
\usepackage[round,authoryear]{natbib}
\usepackage{float}
\usepackage{caption}
\usepackage{changepage}
\usepackage{authblk}

\newcommand{\datasetname}{OrthoFrac-XR}
\newcommand{\best}[1]{\textbf{#1}}
\newcommand{\meanstd}[2]{#1 $\pm$ #2}

\newcolumntype{Y}{>{\centering\arraybackslash}X}

\title{Reliability- and Anatomy-Consistency-Aware Multimodal Learning for Robust Fracture Classification from Bangladeshi Radiographs}

\author[1]{Musa Tur Farazi\thanks{Correspondence: musatur330@gmail.com}}
\author[2]{K G Subarno Bithi}
\affil[1]{Department of Computer Science and Engineering, Bangladesh University of Engineering and Technology, Dhaka 1000, Bangladesh}
\affil[2]{Department of Materials and Metallurgical Engineering, Bangladesh University of Engineering and Technology, Dhaka 1000, Bangladesh}
\date{}

\begin{document}

\maketitle

\begin{abstract}
\noindent Background: Multimodal fracture classifiers may benefit from patient and anatomical metadata, but they can also become brittle when contextual information is missing or mismatched. Methods: We studied 1493 radiographs from the Bangladeshi \datasetname{} dataset using leakage-safe age, sex, bone type, and laterality. A ConvNeXt image encoder was combined with a clinical multilayer perceptron through concatenation, late fusion, reliability-gated residual fusion, and a hierarchical state--location formulation. We additionally introduced an anatomy-consistency gate that attenuates metadata corrections when an image-side anatomical prediction disagrees with the reported bone type. Results: Across five folds and three seeds, hierarchical residual fusion achieved a macro-F1 of $0.6046\pm0.0279$, compared with $0.5727\pm0.0270$ for image-only learning, while improving the Brier score from $0.5239$ to $0.4948$. In a five-fold robustness experiment, anatomy-consistency fusion reduced the macro-F1 loss under shuffled metadata from 0.0567 to 0.0203 relative to ordinary residual fusion, although its clean-data macro-F1 was lower. Without bone type at inference, auxiliary anatomy supervision improved macro-F1 from $0.5620\pm0.0330$ to $0.5899\pm0.0289$. Conclusions: Structured context improves fracture classification, and consistency-aware gating limits harm from mismatched metadata. The observed clean-performance--robustness trade-off and the absence of patient-level identifiers motivate external and prospective validation.

\end{abstract}

\section{Introduction}

Fracture assessment is among the most frequent applications of musculoskeletal radiography. Automated analysis may support triage, prioritization, and second-reader workflows, particularly in settings where specialist availability is limited. Public datasets such as MURA have encouraged the development of image-based musculoskeletal classifiers at scale \citep{rajpurkar2017mura}. More recently, Bangladesh-specific resources have become available: FracAtlas provides classification, localization, and segmentation annotations for musculoskeletal radiographs \citep{abedeen2023fracatlas}, whereas \datasetname{} provides fracture-state and fracture-location labels together with structured patient and anatomical metadata \citep{tabib2026orthofrac}. These resources create an opportunity to study locally relevant models rather than relying exclusively on cohorts collected in high-income health systems.

Image-only models, however, do not reflect the complete information available in routine care. Age, anatomical site, laterality, injury mechanism, and other pre-imaging context can alter disease prevalence and interpretation. Multimodal learning has therefore been explored across radiology and clinical prediction, including the fusion of chest radiographs with electronic health records \citep{hayat2022medfuse}, adaptive patching for paired image--clinical data \citep{zhang2025medpatch}, and image--metadata fusion for infectious and neoplastic disease classification \citep{wu2023deepcovidfuse,xu2016multimodal}. The central promise is that complementary modalities can improve discrimination and calibration when one source alone is insufficient.

This promise is accompanied by a safety concern. A model can learn contextual shortcuts, over-weight a stronger or easier modality, and fail when metadata are missing, corrupted, or assigned to the wrong patient. Missing-modality training and modality dropout partially address absent inputs \citep{neverova2016moddrop,wang2025missing}, but an ``available'' metadata vector is not necessarily a correct metadata vector. For example, an incorrect bone-type field may be syntactically complete while contradicting the anatomy visible in the image. A reliability mechanism that only measures completeness cannot distinguish these cases.

The present study investigates robust multimodal fracture classification using Bangladeshi radiographs. We first establish whether leakage-safe structured metadata improves a strong ConvNeXt image baseline \citep{liu2022convnext}. We then evaluate a reliability-gated residual formulation in which the image prediction remains primary and metadata supplies a learned correction. Because the original four labels mix clinical state (non-fracture and post-fracture) with fracture location (distal and proximal), we also assess a hierarchical state--location formulation. Finally, we introduce an anatomy-consistency gate: an image-side auxiliary head predicts bone type, and the probability assigned to the reported anatomy modulates the metadata correction.

The contributions of this work are as follows:
\begin{enumerate}
    \item A leakage-aware multimodal benchmark on 1493 Bangladeshi radiographs, including five-fold and multi-seed evaluation, probability calibration, metadata-only ablations, and subgroup analysis.
    \item A reliability-gated residual architecture that preserves image-based prediction while allowing structured context to provide class-specific corrections.
    \item A hierarchical formulation that separates fracture state from acute-fracture location and composes both outputs into the original four-class task.
    \item An anatomy-consistency extension and controlled missing, shuffled, and incorrect-metadata experiments that quantify the trade-off between clean accuracy and robustness.
\end{enumerate}

Our results show that multimodal learning provides a reproducible gain over image-only learning. The anatomy-consistency mechanism does not improve clean discrimination over the strongest simple fusion baseline, but it materially reduces degradation under mismatched metadata. We therefore frame the method as a robustness-oriented extension rather than as an unconditional accuracy improvement.

\section{Materials and Methods}

\subsection{Study Design and Reporting}

We conducted a retrospective secondary analysis of the public \datasetname{} dataset. Model development followed a predefined sequence: (i) leakage-safe metadata auditing; (ii) image-only and metadata-only baselines; (iii) standard and residual multimodal fusion; (iv) hierarchical task decomposition; and (v) robustness analysis under controlled metadata corruption. We report discrimination, class-balanced performance, calibration, subgroup behavior, and uncertainty across folds and random seeds. The manuscript was organized with reference to the CLAIM checklist for medical imaging artificial intelligence \citep{mongan2020claim} and TRIPOD+AI reporting principles \citep{collins2024tripodai}; this retrospective benchmark is not presented as a clinical deployment study.

\subsection{Dataset and Cohort}

\datasetname{} contains 1493 orthopedic radiographs collected from four hospitals in Bangladesh and was released with linked structured metadata \citep{tabib2026orthofrac}. The public metadata table contains age, sex, bone type, laterality, bone width, fracture gap, fracture-gap visibility, fracture type, and a primary radiological observation. The four target classes are distal fracture, proximal fracture, post-fracture, and non-fracture. Table~\ref{tab:cohort} summarizes the cohort used in this study.

\begin{table}[H]
\caption{Distribution of the four-class fracture target.\label{tab:cohort}}
\centering
\begin{tabularx}{\textwidth}{lYY}
\toprule
\textbf{Class} & \textbf{Number of Radiographs} & \textbf{Percentage (\%)} \\
\midrule
Non-fracture & 576 & 38.6 \\
Post-fracture & 349 & 23.4 \\
Distal fracture & 314 & 21.0 \\
Proximal fracture & 254 & 17.0 \\
\midrule
Total & 1493 & 100.0 \\
\bottomrule
\end{tabularx}
\end{table}

The public release does not expose a patient identifier. To prevent byte-identical images from crossing folds, each image was hashed with SHA-1 and exact duplicates were assigned to the same group. Four duplicate groups containing eight images were detected. This procedure does not replace patient-level grouping, and the resulting limitation is considered explicitly in Section~\ref{sec:limitations}.

\subsection{Leakage-Safe Clinical Inputs}

The intended prediction is the fracture class from an image and information that could plausibly be available before or independently of radiographic interpretation. We therefore used only age, sex, bone type, and laterality as structured inputs. Bone width, fracture gap, gap visibility, primary observation, and the fracture label were excluded. These variables are measurements or descriptions derived from the radiograph and would create direct or near-direct target leakage.

Age was converted from string to numeric form, median-imputed from the training partition, standardized, and accompanied by a missingness indicator. Categorical fields were one-hot encoded using training-fold vocabularies, with unseen or absent values mapped safely. The fraction of observed metadata fields was retained as an availability scalar $a\in[0,1]$.

We quantified the marginal association between each structured variable and the target using normalized mutual information. Bone type had the strongest association among permitted inputs, followed by age; sex and laterality were weak individually. To identify anatomy shortcuts, metadata-only models were evaluated using five feature sets: sex only, bone type only, age plus sex, age plus sex plus laterality (``safe no anatomy''), and all permitted metadata.

\subsection{Data Partitioning}

Five stratified folds were created while respecting exact-image hash groups. For test fold $f$, fold $(f+1)\bmod 5$ served as validation data and the remaining three folds formed the training set. This design produced approximately 895 training, 299 validation, and 299 test images per split. Main experiments used three random seeds (2026, 2027, and 2028), yielding 15 fold--seed evaluations per model. The targeted anatomy-consistency extension was evaluated across all five folds with seed 2026 because it was introduced after the full multi-seed experiment; this difference in experimental depth is stated throughout the Results.

\subsection{Image and Clinical Encoders}

Radiographs were converted to three channels and resized to $224\times224$ pixels. Training augmentation included random resized cropping, rotation within $\pm7^{\circ}$, and random autocontrast. Images were normalized using ImageNet statistics \citep{deng2009imagenet}. The visual encoder was an ImageNet-pretrained ConvNeXt-Tiny \citep{liu2022convnext}. Let
\begin{equation}
    \mathbf{h}_{I}=E_I(I)\in\mathbb{R}^{d_I}
\end{equation}
represent the image feature vector. Structured inputs $C$ were transformed by a two-layer multilayer perceptron with layer normalization, GELU activations, and dropout:
\begin{equation}
    \mathbf{h}_{C}=E_C(C)\in\mathbb{R}^{d_C}.
\end{equation}
Both representations were projected to a common 256-dimensional fusion space.

\subsection{Baselines}

We compared the following models:
\begin{itemize}
    \item \textbf{Image only}: ConvNeXt-Tiny followed by a four-class linear head.
    \item \textbf{Clinical MLP}: structured metadata without the radiograph.
    \item \textbf{Concatenation}: $[\mathbf{h}_I;\mathbf{h}_C;a]$ followed by a nonlinear classifier.
    \item \textbf{Late fusion}: a validation-selected convex combination of image-only and clinical-only probabilities.
    \item \textbf{Residual Fusion}: the proposed reliability-gated residual formulation.
    \item \textbf{Hierarchical image only and hierarchical residual fusion}: task-decomposed versions described below.
\end{itemize}

Logistic regression and histogram gradient boosting were additionally used for the metadata feature ablation. These classical baselines help establish how much of the target can be inferred from structured context alone.

\subsection{Reliability-Gated Residual Fusion}

Rather than treating image and metadata as equally authoritative, Residual Fusion uses the image prediction as a default and learns a contextual correction. Image and clinical logits are
\begin{equation}
    \mathbf{z}_I=W_I\mathbf{h}_I, \qquad \mathbf{z}_C=W_C\mathbf{h}_C.
\end{equation}
A residual network predicts a class-specific correction
\begin{equation}
    \Delta\mathbf{z}_C=D([\mathbf{h}_I;\mathbf{h}_C]),
\end{equation}
and a reliability network predicts
\begin{equation}
    \mathbf{r}=\sigma\left(G([\mathbf{h}_I;\mathbf{h}_C;a])\right).
\end{equation}
The final logits are
\begin{equation}
    \mathbf{z}=\mathbf{z}_I+a\,\mathbf{r}\odot\Delta\mathbf{z}_C,
    \label{eq:residual}
\end{equation}
where $\odot$ denotes element-wise multiplication. If all metadata are unavailable, $a=0$ and the model falls back to the image branch. Clinical modality dropout was applied during training with probability 0.25.

\subsection{Hierarchical State--Location Prediction}

The four target labels are semantically heterogeneous. Distal and proximal indicate acute-fracture location, whereas non-fracture and post-fracture describe clinical state. We therefore defined
\begin{equation}
 y_{\mathrm{state}}\in\{\text{non-fracture},\text{acute fracture},\text{post-fracture}\}
\end{equation}
and, for acute fractures only,
\begin{equation}
 y_{\mathrm{loc}}\in\{\text{distal},\text{proximal}\}.
\end{equation}
If $p_s$ denotes the state probabilities and $p_l$ denotes location probabilities, the original four-class distribution was reconstructed as
\begin{align}
 p(\text{non-fracture}) &= p_s(\text{non-fracture}),\\
 p(\text{post-fracture}) &= p_s(\text{post-fracture}),\\
 p(\text{distal}) &= p_s(\text{acute})p_l(\text{distal}),\\
 p(\text{proximal}) &= p_s(\text{acute})p_l(\text{proximal}).
\end{align}
The location loss was applied only to acute-fracture examples.

\subsection{Anatomy-Consistency-Gated Fusion}

Residual reliability in Equation~\eqref{eq:residual} measures learned usefulness and input availability, but it does not explicitly detect a semantically wrong bone-type field. We introduced an image-side anatomy head
\begin{equation}
    \mathbf{q}=\operatorname{softmax}(W_b\mathbf{h}_I),
\end{equation}
where $q_b$ is the predicted probability of bone type $b$. Given a reported bone type $b_C$, the anatomy-consistency score is
\begin{equation}
 c(I,C)=\alpha+(1-\alpha)q_{b_C}^{\gamma},
 \label{eq:consistency}
\end{equation}
with floor $\alpha=0.05$ and exponent $\gamma=1$. When bone type is unavailable, $c=1$ so that age, sex, and laterality are not suppressed. The consistency-aware prediction is
\begin{equation}
    \mathbf{z}=\mathbf{z}_I+a\,c(I,C)\,\mathbf{r}\odot\Delta\mathbf{z}_C.
    \label{eq:consistencyfusion}
\end{equation}

\begin{figure}[H]
\centering
\includegraphics[width=0.98\textwidth]{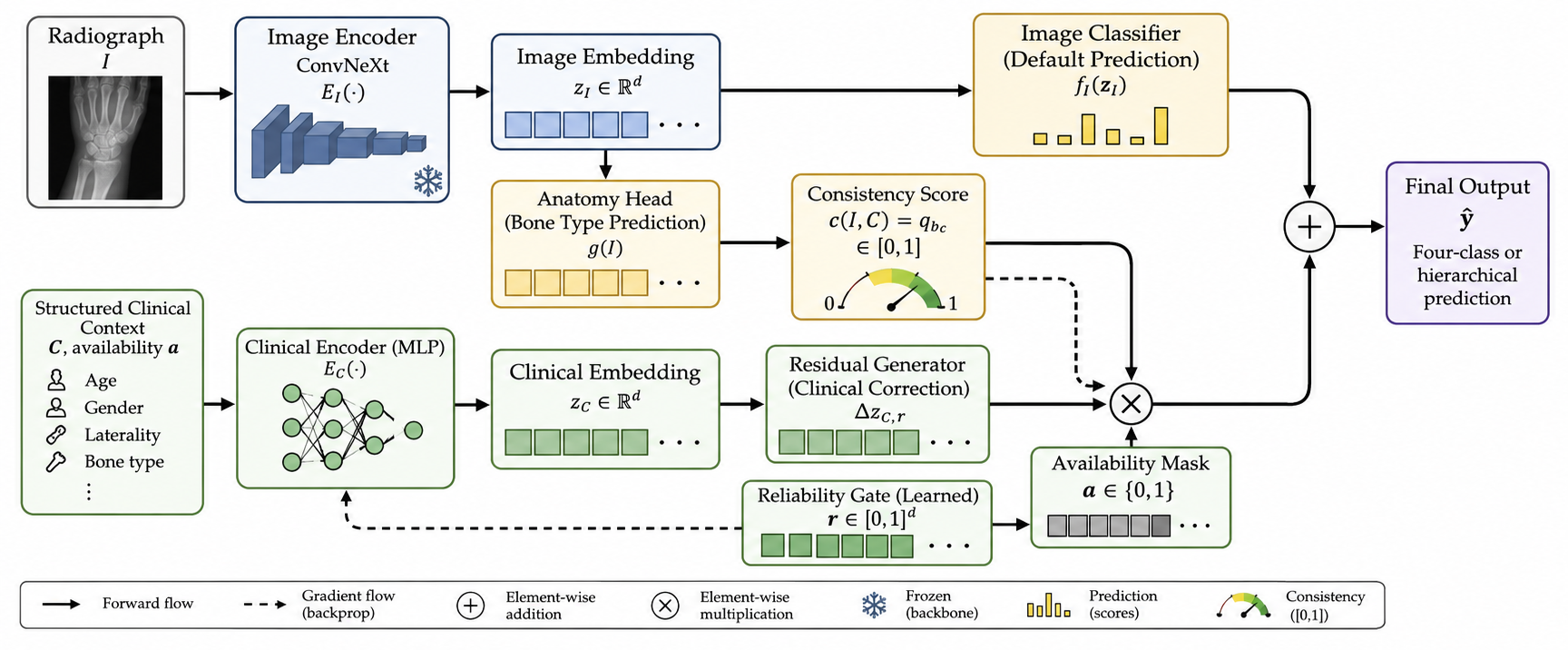}
\caption{Overview of the reliability- and anatomy-consistency-aware residual fusion architecture.\label{fig:architecture}}
\end{figure}

Figure~\ref{fig:architecture} summarizes the architecture. The same mechanism was evaluated in flat and hierarchical variants.

\subsection{Training Objectives}

Class-weighted cross-entropy with label smoothing of 0.05 supervised the four-class output. Auxiliary image and clinical heads were weighted by 0.25 and 0.05, respectively. Hierarchical models additionally used state and acute-location losses. The anatomy head used a 17-class bone-type cross-entropy loss with weight 0.25.

To avoid the earlier failure mode in which shuffled metadata were simply trained with the original class label, we used a counterfactual ranking loss. For the true-class score $s_y(I,C)$ and shuffled context $C'$,
\begin{equation}
 \mathcal{L}_{\mathrm{rank}}=\max\left(0,m-s_y(I,C)+s_y(I,C')\right),
\end{equation}
where $m=0.20$. The anatomy-consistency models additionally used a safe-fallback objective under shuffled context:
\begin{equation}
 \mathcal{L}_{\mathrm{fallback}}=D_{\mathrm{KL}}\left(p(I,C')\;\|\;\operatorname{stopgrad}p_I(I)\right).
\end{equation}
The total objective was a weighted sum of classification, auxiliary, hierarchical, ranking, anatomy, and fallback terms.

Models were optimized using AdamW with backbone learning rate $10^{-5}$, head learning rate $3\times10^{-4}$, weight decay $10^{-4}$, batch size 16, cosine learning-rate decay, mixed precision, and early stopping based on validation macro-F1. The main experiment allowed up to 18 epochs with patience 4.

\subsection{Robustness Conditions}

We evaluated metadata robustness by modifying only the test metadata while leaving the image unchanged:
\begin{itemize}
    \item random missingness at 25\%, 50\%, and 75\% per field;
    \item complete metadata removal;
    \item shuffling all structured fields across patients;
    \item replacement of bone type for 50\% or 100\% of examples with a different valid category.
\end{itemize}
Each stochastic condition was repeated ten times in the five-fold targeted experiment. Validation-fitted temperature scaling was applied to both clean and corrupted predictions so that Brier score and expected calibration error (ECE) were comparable.

\subsection{Evaluation and Statistical Analysis}

The primary metric was macro-F1 because the four classes were imbalanced and clinically distinct. We additionally report accuracy, balanced accuracy, one-vs-rest macro-AUROC, negative log-likelihood, multiclass Brier score \citep{brier1950}, and ECE \citep{guo2017calibration}. Temperature scaling was fitted on each validation fold. For the main five-fold, three-seed analysis, results are reported as the mean and standard deviation over 15 runs. Paired differences were evaluated on matched fold--seed results using the Wilcoxon signed-rank test and by prediction-level paired permutation analysis. For the one-seed anatomy-consistency experiment, paired bootstrap intervals were calculated from out-of-fold predictions; these intervals quantify sample uncertainty but not random-seed uncertainty.

Grad-CAM \citep{selvaraju2017gradcam} was used for qualitative inspection. Because \datasetname{} does not provide lesion bounding boxes, heatmaps were not interpreted as quantitative localization evidence.

\section{Results}

\subsection{Metadata Association and Metadata-Only Performance}

The primary radiological observation had normalized mutual information of 0.770 with the target and gap visibility had 0.211; both were excluded because they encode post-imaging evidence. Among permitted inputs, bone type had the largest association (0.113), followed by age (0.062), whereas laterality and sex were weak (below 0.01).

\begin{figure}[H]
\centering
\includegraphics[width=0.90\textwidth]{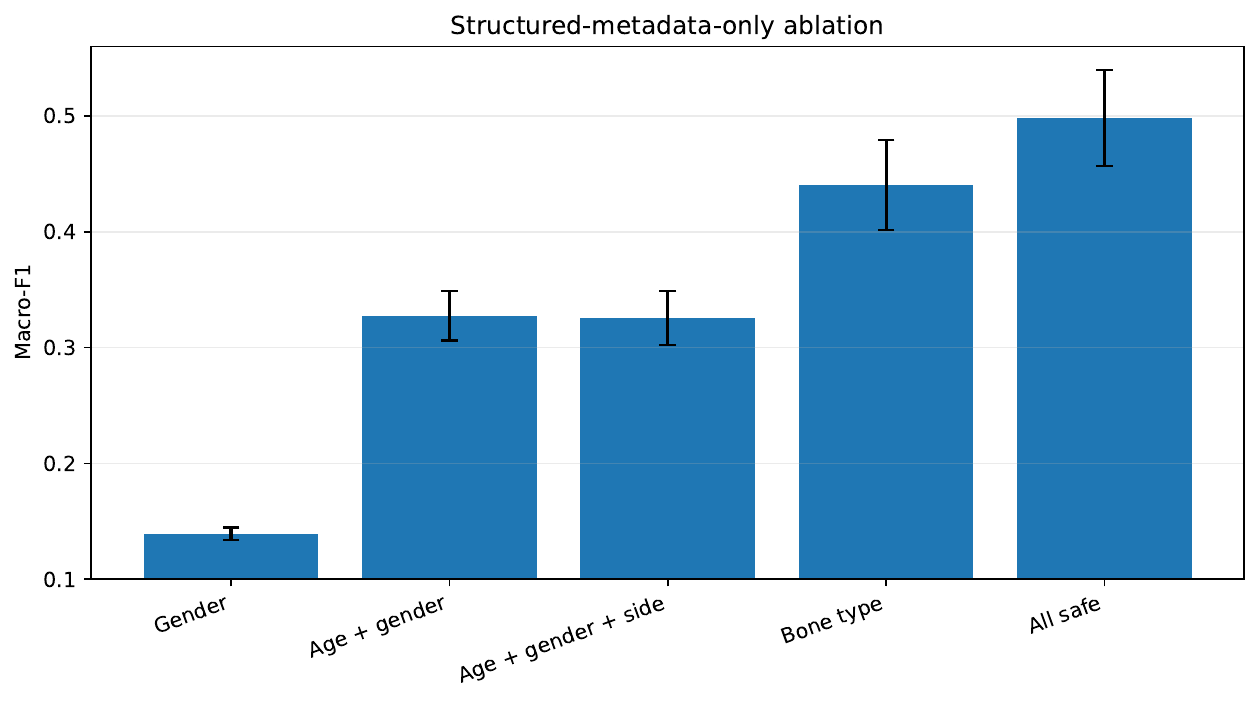}
\caption{Metadata-only feature ablation across five folds. Bars show mean macro-F1; error bars show the standard deviation across folds.\label{fig:metadata}}
\end{figure}

Figure~\ref{fig:metadata} and Table~\ref{tab:metadata} show that metadata alone contained substantial class information. Histogram gradient boosting using all permitted metadata achieved a macro-F1 of $0.4984\pm0.0413$. Bone type alone achieved $0.4401\pm0.0388$, while sex alone was near chance for a four-class macro metric. The gain from all metadata over the no-anatomy feature set indicates that anatomical site is the dominant structured signal and a potential shortcut that must be audited.

\begin{table}[H]
\caption{Best metadata-only results for each feature set using histogram gradient boosting (five folds).\label{tab:metadata}}
\centering
\begin{tabularx}{\textwidth}{lYYY}
\toprule
\textbf{Feature Set} & \textbf{Macro-F1} & \textbf{Balanced Accuracy} & \textbf{Macro-AUROC} \\
\midrule
Sex only & \meanstd{0.1391}{0.0055} & 0.2500 & \meanstd{0.5167}{0.0135} \\
Age + sex & \meanstd{0.3274}{0.0212} & 0.3300 & \meanstd{0.6127}{0.0223} \\
Age + sex + laterality & \meanstd{0.3254}{0.0231} & 0.3270 & \meanstd{0.6146}{0.0219} \\
Bone type only & \meanstd{0.4401}{0.0388} & 0.4510 & \meanstd{0.6873}{0.0220} \\
All permitted metadata & \best{\meanstd{0.4984}{0.0413}} & \best{0.5030} & \best{\meanstd{0.7498}{0.0132}} \\
\bottomrule
\end{tabularx}
\end{table}

\subsection{Main Five-Fold, Three-Seed Comparison}

Table~\ref{tab:main} presents the principal comparison across 15 matched fold--seed runs. Image-only ConvNeXt achieved macro-F1 $0.5727\pm0.0270$. Every multimodal strategy improved the mean macro-F1. Concatenation reached $0.6015\pm0.0234$, residual fusion reached $0.6027\pm0.0214$, and hierarchical residual fusion achieved the highest mean macro-F1 of $0.6046\pm0.0279$.

The hierarchical residual model improved macro-F1 by 0.0319 and Brier score by 0.0291 relative to image-only learning. It was better than image-only in 13 of 15 matched runs; a one-sided Wilcoxon signed-rank test over the fold--seed results gave $p<0.001$. Prediction-level paired permutation tests similarly favored each multimodal model over image only ($p=0.0002$ for macro-F1, balanced accuracy, and Brier score). The hierarchy itself was not responsible for most of the gain: hierarchical image only ($0.5731\pm0.0258$) was nearly identical to ordinary image only. Moreover, the differences among concatenation, residual fusion, and hierarchical residual fusion were small and did not establish a decisive superiority of the more complex architectures.

\begin{table}[H]
\caption{Main performance across five folds and three random seeds. Values are means over 15 runs. Higher is better except for Brier score and ECE.\label{tab:main}}
\centering
\scriptsize
\begin{tabularx}{\textwidth}{lYYYYYY}
\toprule
\textbf{Model} & \textbf{Accuracy} & \textbf{Balanced Acc.} & \textbf{Macro-F1} & \textbf{Macro-AUROC} & \textbf{Brier $\downarrow$} & \textbf{ECE $\downarrow$} \\
\midrule
Clinical MLP & 0.4945 & 0.4978 & 0.4830 & 0.7436 & 0.6444 & 0.0730 \\
Image only & 0.5988 & 0.5735 & 0.5727 & 0.8331 & 0.5239 & 0.0698 \\
Hierarchical image only & 0.6048 & 0.5695 & 0.5731 & 0.8346 & 0.5189 & 0.0671 \\
Late fusion & 0.6115 & 0.5890 & 0.5873 & 0.8433 & 0.5219 & 0.0959 \\
Concatenation & 0.6249 & 0.6009 & 0.6015 & \best{0.8491} & 0.4976 & 0.0702 \\
Residual Fusion & \best{0.6289} & 0.6019 & 0.6027 & 0.8472 & 0.4969 & 0.0603 \\
Hierarchical Residual Fusion & 0.6278 & \best{0.6042} & \best{0.6046} & 0.8488 & \best{0.4948} & \best{0.0600} \\
\bottomrule
\end{tabularx}
\end{table}

\begin{figure}[H]
\centering
\includegraphics[width=0.92\textwidth]{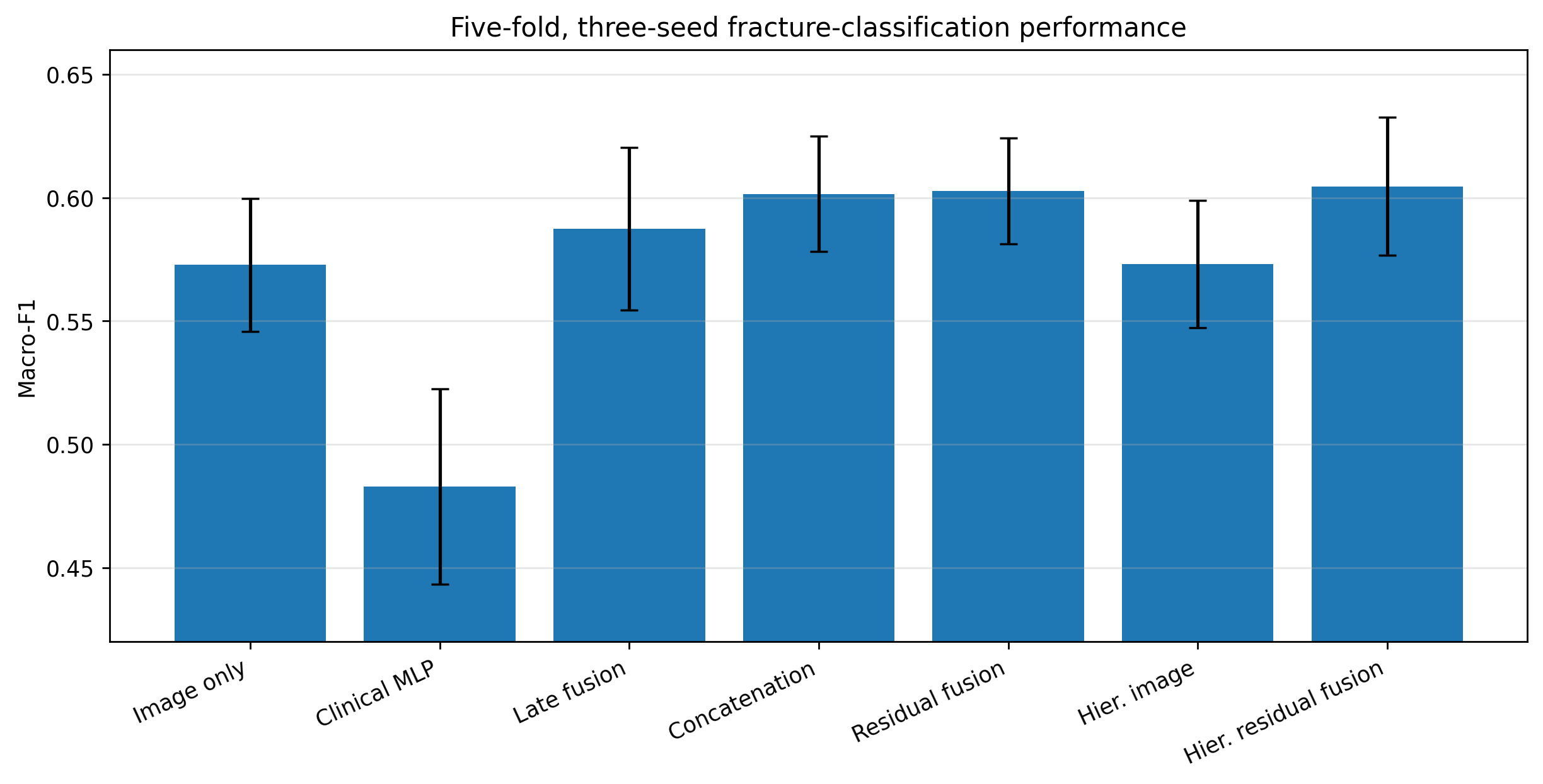}
\caption{Macro-F1 comparison for the main five-fold, three-seed experiment. Error bars indicate standard deviation over 15 runs.\label{fig:mainf1}}
\end{figure}

\subsection{Anatomy-Consistency Extension: Clean Performance}

The anatomy-consistency extension was evaluated in a targeted five-fold, one-seed experiment. Table~\ref{tab:consistencyclean} reports clean performance with all permitted metadata and with bone type removed from the clinical vector. With all metadata, simple concatenation achieved the highest macro-F1 ($0.6031\pm0.0265$). Anatomy-Consistency Fusion achieved $0.5846\pm0.0128$, and its hierarchical version achieved $0.5965\pm0.0418$. Paired bootstrap intervals versus concatenation included zero and were centered below zero for both consistency variants; thus anatomy-consistency fusion did not provide a clean-data accuracy advantage.

When bone type was excluded at inference, however, Anatomy-Consistency Fusion achieved $0.5899\pm0.0289$, compared with $0.5620\pm0.0330$ for ordinary residual fusion. The paired macro-F1 improvement was 0.0262 with a 95\% bootstrap interval of approximately [0.0031, 0.0491]. Since the anatomy head was still supervised during training but no bone field was supplied to the gate at inference, this result indicates that auxiliary anatomical supervision improved the image representation independently of metadata gating.

\begin{table}[H]
\caption{Mean clean performance across five folds in the one-seed anatomy-consistency experiment.\label{tab:consistencyclean}}
\centering
\scriptsize
\begin{tabularx}{\textwidth}{llYYYY}
\toprule
\textbf{Feature Set} & \textbf{Model} & \textbf{Macro-F1} & \textbf{Macro-AUROC} & \textbf{Brier $\downarrow$} & \textbf{ECE $\downarrow$} \\
\midrule
\multirow{6}{*}{All metadata} & Image only & 0.5577 & 0.8225 & 0.5361 & 0.0649 \\
& Concatenation & \best{0.6031} & \best{0.8503} & \best{0.4934} & 0.0703 \\
& Residual Fusion & 0.6012 & 0.8480 & 0.4993 & 0.0603 \\
& Hierarchical Residual Fusion & 0.5821 & 0.8476 & 0.5031 & 0.0700 \\
& Anatomy-Consistency Fusion & 0.5846 & 0.8441 & 0.5044 & \best{0.0571} \\
& Hierarchical Anatomy-Consistency Fusion & 0.5965 & 0.8429 & 0.5080 & 0.0578 \\
\midrule
\multirow{4}{*}{No bone metadata} & Concatenation & 0.5763 & 0.8314 & 0.5261 & -- \\
& Residual Fusion & 0.5620 & 0.8252 & 0.5439 & -- \\
& Anatomy-Consistency Fusion & \best{0.5899} & \best{0.8378} & \best{0.5172} & -- \\
& Hierarchical Anatomy-Consistency Fusion & 0.5691 & 0.8334 & -- & -- \\
\bottomrule
\end{tabularx}
\end{table}

\subsection{Robustness to Missing and Mismatched Metadata}

\begin{figure}[H]
\centering
\includegraphics[width=0.92\textwidth]{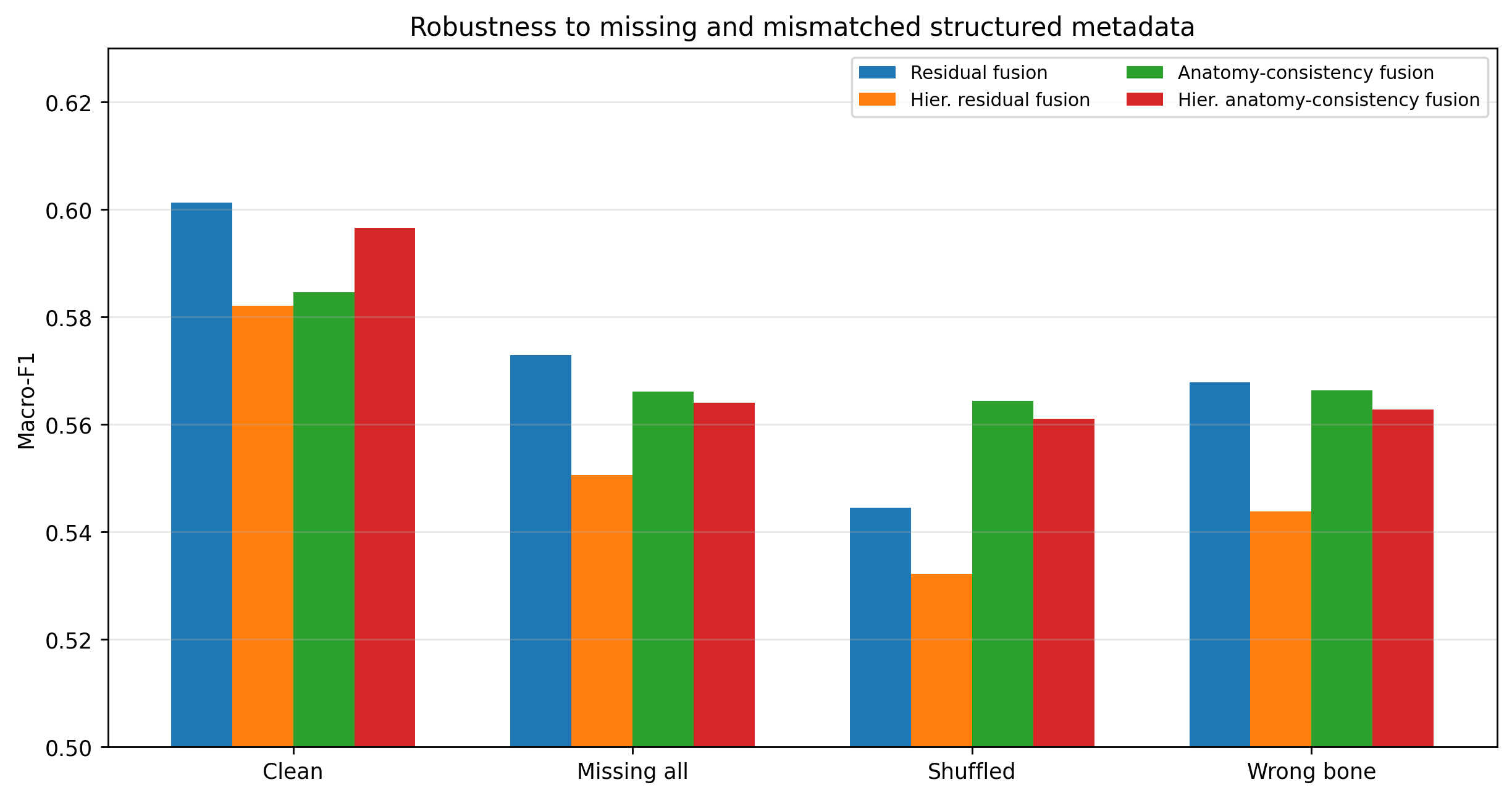}
\caption{Macro-F1 under clean, missing, shuffled, and incorrect metadata in the five-fold targeted robustness experiment.\label{fig:robustness}}
\end{figure}

Figure~\ref{fig:robustness} and Table~\ref{tab:robustness} compare clean and corrupted metadata conditions. The most pronounced effect occurred when all structured fields were shuffled across examples. Ordinary residual fusion decreased from macro-F1 0.6012 to 0.5445, a loss of 0.0567. Anatomy-consistency fusion decreased from 0.5846 to 0.5643, a loss of only 0.0203. The hierarchical consistency model also reduced the shuffled-context loss relative to its hierarchical residual-fusion counterpart (0.0355 versus 0.0500).

Under 100\% wrong bone type, the clean-to-corrupted loss was 0.0334 for residual fusion and 0.0183 for anatomy-consistency fusion. The consistency mechanism therefore achieved its intended safety effect: it suppressed a contradictory contextual correction. Nevertheless, its lower clean starting point meant that absolute corrupted performance was similar to ordinary residual fusion. Missing all metadata caused graceful degradation for every residual model, confirming that the image branch remained functional.

\begin{table}[H]
\caption{Robustness of residual models using all permitted metadata. Values are mean macro-F1 over five folds; stochastic corruptions were repeated ten times per fold. Parentheses show the decrease from each model's own clean performance.\label{tab:robustness}}
\centering
\resizebox{\textwidth}{!}{%
\begin{tabular}{lccccc}
\toprule
\textbf{Model} & \textbf{Clean} & \textbf{All Missing} & \textbf{Shuffled} & \textbf{50\% Wrong Bone} & \textbf{100\% Wrong Bone} \\
\midrule
Residual Fusion & 0.6012 & 0.5728 ($-0.0284$) & 0.5445 ($-0.0567$) & 0.5855 ($-0.0158$) & 0.5678 ($-0.0334$) \\
Hierarchical Residual Fusion & 0.5821 & 0.5505 ($-0.0316$) & 0.5321 ($-0.0500$) & 0.5632 ($-0.0190$) & 0.5438 ($-0.0383$) \\
Anatomy-Consistency Fusion & 0.5846 & 0.5661 ($-0.0185$) & \best{0.5643 ($-0.0203$)} & \best{0.5767 ($-0.0079$)} & \best{0.5663 ($-0.0183$)} \\
Hierarchical Anatomy-Consistency Fusion & 0.5965 & 0.5640 ($-0.0325$) & 0.5610 ($-0.0355$) & 0.5798 ($-0.0166$) & 0.5627 ($-0.0338$) \\
\bottomrule
\end{tabular}%
}
\end{table}

The mean anatomy-consistency probability for correctly matched all-metadata samples ranged from approximately 0.41 to 0.53 across folds. This moderate value explains part of the clean-data trade-off: the multiplicative gate attenuated metadata corrections even when the context was correct. A softer transformation or explicit calibration of the anatomy head may preserve more clean performance.

\subsection{Class-Wise Behavior and Calibration}

\begin{figure}[H]
\centering
\begin{minipage}[t]{0.49\textwidth}
\centering
\includegraphics[width=\textwidth]{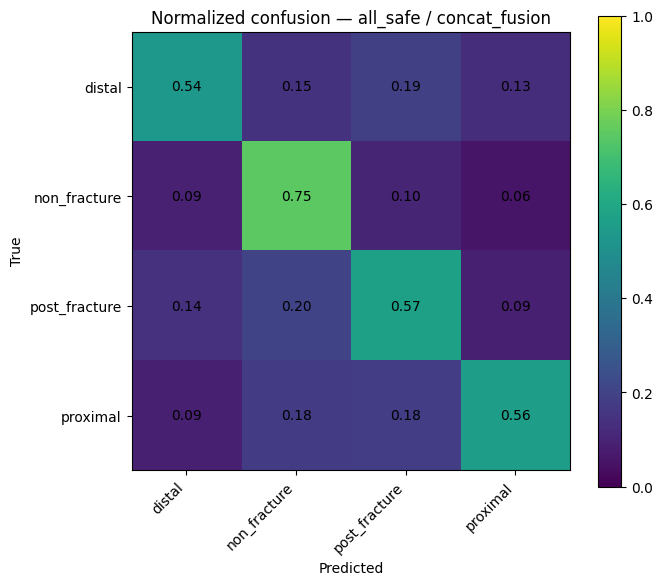}
\end{minipage}\hfill
\begin{minipage}[t]{0.49\textwidth}
\centering
\includegraphics[width=\textwidth]{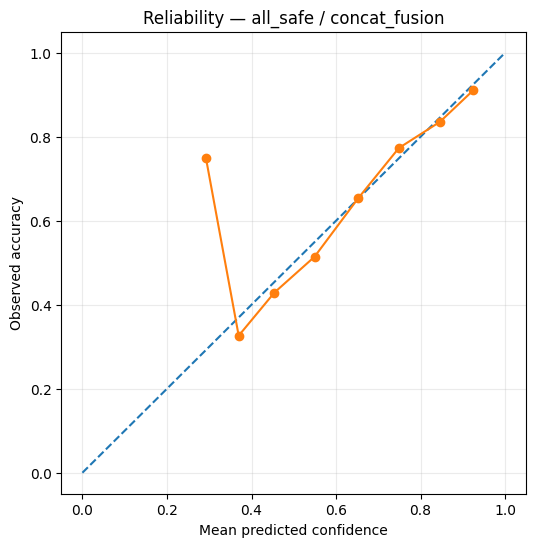}
\end{minipage}
\caption{Representative diagnostics for concatenation in the targeted five-fold experiment: normalized confusion matrix (left) and reliability diagram after validation-fitted temperature scaling (right).\label{fig:diagnostics}}
\end{figure}

In the targeted experiment, concatenation was the strongest clean model. Its class-wise F1 scores were 0.554 for distal fracture, 0.737 for non-fracture, 0.557 for post-fracture, and 0.565 for proximal fracture. Figure~\ref{fig:diagnostics} shows the normalized confusion matrix and reliability diagram. Non-fracture cases were easiest, while post-fracture appearance was confused with both acute-fracture locations. This is consistent with the visual heterogeneity of healed, treated, or previously fractured anatomy and indicates that longitudinal or treatment-history variables could be especially useful.

The full multi-seed experiment showed that residual fusion improved ECE from $0.0698\pm0.0156$ for image only to approximately $0.0600$ for both residual variants. Late fusion had the worst ECE despite improving discrimination, illustrating that accuracy and probability reliability must be evaluated separately.

\subsection{Subgroup Findings}

The cohort was predominantly pediatric: 1054 of 1491 images with valid age (70.6\%) were from patients aged 18 years or younger. Performance was therefore less precisely estimated for older groups, particularly those above 60 years ($n=52$). Sex-specific macro-F1 values were similar, whereas right-sided images outperformed left-sided images by approximately eight percentage points in the targeted analysis. This difference was visible in image-only models as well as multimodal models, suggesting that it may reflect class composition, anatomy, acquisition, or dataset construction rather than the metadata branch alone. Several bone-type subgroups contained small samples or lacked one or more target classes; subgroup AUROC was therefore undefined in some cases and should not be interpreted as evidence of fairness or generalization.

\section{Discussion}

\subsection{Principal Findings}

This study provides three main findings. First, leakage-safe structured context improved fracture classification over image-only learning. Across five folds and three seeds, the strongest multimodal models improved macro-F1 by approximately three percentage points and reduced Brier score, with gains observed in most matched runs. This supports the premise that age and anatomy provide complementary priors even when post-imaging observations are rigorously excluded.

Second, architectural complexity was not automatically beneficial. Concatenation, residual fusion, and hierarchical residual fusion had nearly identical mean performance. Hierarchical image-only learning did not improve over the flat image baseline, indicating that task decomposition was clinically coherent but not independently responsible for the multimodal gain. The appropriate claim is therefore not that hierarchy decisively outperforms standard fusion, but that reliability-gated and hierarchical variants can match strong simple baselines while providing a structured fallback mechanism.

Third, the anatomy-consistency gate reduced sensitivity to context that was complete but wrong. Under shuffled metadata, anatomy-consistency fusion lost 0.0203 macro-F1 compared with 0.0567 for ordinary residual fusion. Under fully wrong bone type, its loss was approximately half that of residual fusion. This behavior is aligned with the design of Equation~\eqref{eq:consistencyfusion}: the image branch remains primary, and contradictory anatomy attenuates the clinical correction. The trade-off was lower clean macro-F1 because the image anatomy probability was only moderate even for correct metadata.

\subsection{Relation to Prior Multimodal Work}

Prior multimodal medical systems have shown that imaging and structured records can be complementary \citep{hayat2022medfuse,zhang2025medpatch,wu2023deepcovidfuse}. Our findings agree with that literature but emphasize a distinction between missing and mismatched modalities. Modality dropout \citep{neverova2016moddrop} and missing-enabled fusion \citep{wang2025missing} train a model to tolerate absent inputs. In contrast, a wrong bone type is present and well-formed; an availability gate alone treats it as trustworthy. Anatomy-consistency gating introduces a cross-modal semantic check that is specific to the visible anatomy.

The no-anatomy experiment further suggests that an auxiliary anatomical task can regularize the image encoder. This is consistent with multi-task representation learning: predicting the anatomical category encourages the visual branch to organize features by skeletal structure, even when no bone-type field is supplied at inference. The significant improvement of anatomy-consistency fusion over residual fusion in this setting may thus be driven more by auxiliary supervision than by the multiplicative gate. Future ablations should separate the anatomy head, gate, ranking loss, and fallback loss.

\subsection{Clinical and Regional Relevance}

The use of a Bangladesh-specific cohort is valuable because imaging distributions, equipment, referral patterns, and patient age profiles can differ across health systems. Nevertheless, ``regional'' does not imply representative of all Bangladeshi hospitals. \datasetname{} is a retrospective public dataset from four institutions, and the cohort is heavily pediatric. The model should be regarded as a regional benchmark rather than a validated national diagnostic tool.

The task also differs from the broader symptom--image disease-detection objective. The present dataset provides structured demographics and anatomy, not genuine chief complaints or symptom narratives. Accordingly, this manuscript describes multimodal image--metadata classification and does not claim symptom-grounded reasoning. A future local cohort should include pre-imaging pain location, mechanism of injury, swelling, functional limitation, time since injury, prior fracture history, and treatment status.

\subsection{Implications of the Clean--Robustness Trade-Off}

The strongest clean model in the targeted experiment was concatenation, whereas the consistency gate was safer under incorrect context. In clinical machine learning, this trade-off should be made explicit rather than hidden by reporting one aggregate score. If metadata correctness is highly reliable, concatenation may be preferred for clean discrimination. If records may be mismatched, partially synchronized, or manually entered, the consistency-aware model provides a more conservative failure mode.

Several modifications may improve the trade-off. First, the consistency transformation can use a higher floor, such as $c'=0.5+0.5q_{b_C}$, to avoid halving correct residuals. Second, the bone head can be calibrated separately or trained with class-balanced loss. Third, an explicit binary match/mismatch discriminator can be supervised with synthetic swaps. Fourth, consistency can be applied to anatomy-dependent portions of the clinical representation rather than to age and sex corrections globally. These directions are prospective and were not evaluated in the present study.

\subsection{Limitations}\label{sec:limitations}

This study has several important limitations.
\begin{enumerate}
    \item \textbf{No patient-level identifier.} Exact duplicates were grouped, but multiple non-identical radiographs from the same patient may still occur across folds. This can only be resolved by obtaining patient or study identifiers from the data custodians or by validating on a separate cohort.
    \item \textbf{No external validation.} All training and testing used only \datasetname{}. Performance across hospitals, devices, adult populations, and other countries remains unknown.
    \item \textbf{One-seed anatomy-consistency evaluation.} The main residual-fusion experiment used three seeds, but the anatomy-consistency extension used one seed across five folds. Its clean and robustness effects require multi-seed confirmation.
    \item \textbf{Structured metadata, not symptoms.} The model used age, sex, anatomy, and laterality. It did not use chief complaints, injury mechanism, or free-text symptoms.
    \item \textbf{Anatomical shortcut.} Bone type was the strongest metadata-only predictor. Although clinically relevant, it can encode class prevalence and inflate multimodal gains. We therefore reported a no-anatomy experiment, but further cross-anatomy and site-held-out validation is needed.
    \item \textbf{Retrospective labels.} The four classes mix state and location and may not fully capture fracture complexity, severity, treatment, or uncertainty. No independent re-reading by a blinded radiologist was performed for this study.
    \item \textbf{Qualitative explainability.} Grad-CAM was used only for visualization; without lesion boxes in \datasetname{}, localization accuracy was not quantified.
\end{enumerate}

\section{Conclusions}

Leakage-safe structured metadata improved fracture classification from Bangladeshi radiographs over a strong image-only ConvNeXt baseline. Reliability-gated residual and hierarchical fusion achieved the best mean class-balanced and calibration results in the full five-fold, three-seed experiment, although their clean discrimination was statistically close to simple concatenation. An image-derived anatomy-consistency gate reduced degradation under shuffled and incorrect metadata, demonstrating a safer fallback behavior at the cost of some clean accuracy. Auxiliary anatomical supervision also improved performance when bone type was unavailable at inference. These results support robustness-aware multimodal learning as a promising research direction, but patient-level, external, and prospective validation is required before clinical use.

\section*{Author Contributions}
Conceptualization, M.T.F., K.G.S.B.; methodology, M.T.F.; software, M.T.F., K.G.S.B; validation, M.T.F., K.G.S.B; formal analysis, M.T.F.; investigation, M.T.F.; data curation, M.T.F., K.G.S.B; writing---original draft preparation, M.T.F.; writing---review and editing, M.T.F., K.G.S.B; visualization, M.T.F., K.G.S.B; All authors have read and agreed to the published version of the manuscript.

\section*{Funding}
This research received no external funding.

\section*{Institutional Review Board Statement}
This study was a secondary analysis of a publicly released, de-identified dataset and involved no new patient recruitment or intervention. The original data collection was approved by the relevant research ethics committee under protocol REC-FSIT-2025/12639, as reported by the dataset authors.

\section*{Informed Consent Statement}
Patient consent procedures and de-identification were managed by the original dataset creators. No identifiable information was accessed in the present secondary analysis.

\section*{Data Availability Statement}
The \datasetname{} dataset is publicly available through Figshare at \url{https://doi.org/10.6084/m9.figshare.32021085.v1}. The training notebooks, fold definitions, predictions, and analysis tables generated for this study are provided as Supplementary Materials with this submission.

\section*{Acknowledgments}
The authors acknowledge the creators and contributing hospitals of \datasetname{} for releasing the radiographs and metadata used in this study. During preparation of this manuscript, the authors used OpenAI ChatGPT for manuscript structuring, language refinement, LaTeX formatting, and assistance in checking analysis code. The authors reviewed and edited all generated material and take full responsibility for the content of this publication.

\section*{Conflicts of Interest}
The authors declare no conflicts of interest.

\section*{Abbreviations}
The following abbreviations are used in this manuscript:\\
\noindent
\begin{tabular}{@{}ll}
AI & Artificial intelligence\\
AUROC & Area under the receiver operating characteristic curve\\
ECE & Expected calibration error\\
KL & Kullback--Leibler\\
MLP & Multilayer perceptron
\end{tabular}

\end{document}